# Progress toward high-Q perfect absorption: A Fano antilaser

Sunkyu Yu[†], Xianji Piao[†], Jiho Hong, and Namkyoo Park[*]

*Photonic Systems Laboratory, School of EECS, Seoul National University, Seoul 151-744, Korea*

Here we propose a route to the high-$Q$ perfect absorption of light by introducing the concept of a Fano anti-laser. Based on the drastic spectral variation of the optical phase in a Fano-resonant system, a spectral singularity for scatter-free perfect absorption can be achieved with an order of magnitude smaller material loss. By applying temporal coupled mode theory to a Fano-resonant waveguide platform, we reveal that the required material loss and following absorption $Q$-factor are ultimately determined by the degree of Fano spectral asymmetry. The feasibility of the Fano anti-laser is confirmed using a photonic crystal platform, to demonstrate spatio-spectrally selective heating. Our results utilizing the phase-dependent control of device bandwidths derive a counterintuitive realization of high-$Q$ perfect conversion of light into internal energy, and thus pave the way for a new regime of absorption-based devices, including switches, sensors, thermal imaging, and opto-thermal emitters.

Light absorption is a fundamental phenomenon of the conversion of electromagnetic energy into internal energies of different forms, and has been exploited in various energy conversion devices [1-6]. Unconventional physics have recently been derived from the use of optical loss as well, *e.g.,* parity-time symmetric optics [7-10] and anomalous refraction [11]. The quest toward *ultimate* optical absorption has also progressed under the notion of 'perfect absorption' [12-22], which utilizes the destructive interference between scatterings from different origins. Employing the concept of coherent perfect absorption (CPA) [23-30], the origin of perfect absorption has been explained intuitively in the context of a time-reversed laser at the lasing threshold, *i.e.,* an *anti-laser* with a scattering singularity.

The applications of optical absorption can be classified in terms of their spectral responses: broadband absorptions [18-20] for photovoltaics [5], anti-reflector, and band-stop filter, and narrowband ones [12-17] for high-$Q$ sensor [1-3,12,13], thermal imaging [14], opto-thermal emitter [4,15,16], and absorptive switching [6]. Interestingly, although a clear description of perfect absorption can be made through the concept of an anti-laser [23], efforts to develop narrowband anti-laser have been hindered by the difficulties in achieving high-$Q$ values; in stark contrast to a laser which supports a narrower, single-wavelength spectrum *above* the threshold. This restriction toward high-$Q$ anti-laser is a result of the fundamental tradeoff between the absorption and $Q$-factor in the system. Because the loaded $Q$-factor $Q_L$ is related to coupling $Q_C$ and intrinsic $Q_I$ (including material loss) through $1/Q_L = 1/Q_I + 1/Q_C$, it is impossible to achieve both large absorption ($Q_I\downarrow$) and a narrow spectrum ($Q_L\uparrow$) simultaneously even with the infinite $Q_C$ of isolated absorbers, hindering the realization of narrowband anti-lasing *above* the threshold. Consequently, past approaches for narrowband perfect absorber based on a single mode [12-17] accompany inherent technical bottleneck, most critically, from the tradeoff between 'large absorption' and 'narrow bandwidth'. Considering important applications of a high-$Q$ perfect absorber [12-17], the obstacles for high-$Q$ anti-lasing should be overcome, enabling progress toward a 'single-wavelength absorber'.

In this paper, we propose a novel path toward high-$Q$ perfect absorption by employing the time-reversed form of a Fano laser [31], *i.e.,* a Fano anti-laser. By exploiting the drastic phase evolution inside Fano-resonant systems [32-35], the cancellation condition of scattering for CPA changes rapidly, deriving an order of increasing $Q$ for an absorption spectrum: leading to the perfect absorption with smaller material loss ($Q_I\uparrow$). Using coupled mode theory (CMT) [36], we achieve physical quantities for the control of the CPA bandwidth, determined by the degree of the famous Fano spectral asymmetry [32-35]. The implementation of photonic crystal (PC) Fano anti-laser is also presented, with heat generation maps for applications in spatio-spectrally selective thermal emitters and absorptive switching. With its drastic manipulation of absorption and its structural simplicity based on a two-level design, we demonstrate that the Fano anti-laser forms a suitable platform for high-$Q$ conversion of light into internal energy.

First, we consider the underlying physics of CPA [23] in terms of the scattering matrix (S-matrix). Considering a black-box potential connected to two ports, 1 and 2, its frequency-dependent S-matrix relation is straightforward:

$$\begin{bmatrix} S_{-1} \\ S_{-2} \end{bmatrix} = \begin{bmatrix} r_{11}(\omega) & t(\omega) \\ t(\omega) & r_{22}(\omega) \end{bmatrix} \begin{bmatrix} S_{+1} \\ S_{+2} \end{bmatrix}. \quad (1)$$

where $S_+$ (or $S_-$) is the incoming (or outgoing) wave amplitude for each port and $r(\omega)$ and $t(\omega)$ denote the reflection and transmission coefficients, respectively. To achieve perfect absorption ($S_{-1} = S_{-2} = 0$ for nonzero $S_{+1}$ or $S_{+2}$), the S-matrix should be *singular* as $D(\omega) = r_{11}(\omega)\cdot r_{22}(\omega) - t^2(\omega) = 0$. Therefore, the bandwidth and Q-factor of perfect absorption are determined by the robustness of the condition $D(\omega) \sim 0$. In stark contrast to the cases of wideband perfect absorption [18-20], high-Q perfect absorption thus requires a drastic spectral change of the singularity parameter $D(\omega)$. To satisfy this condition even in the presence of material loss which degrades $Q_l$, we employ Fano interference, which derives phase-induced spectral asymmetry [32-35].

Considering the potential applications of high-Q perfect absorption including narrowband absorptive switching or filtering, here we focus on the guided-wave platform, in contrast to other metamaterial perfect absorbers [12-22]. Figure 1a shows the simplest design of the Fano anti-laser for guided-waves, supporting the time-reversed wave dynamics of lasing in a Fano system [31]. This structure can be realized with e.g. the use of a two-port waveguide connected to a narrowband stub (or ring) structure through a broadband junction resonator [34,35]. The stub supports a complex refractive index $n$, where $Im[n] > 0$ for a Fano laser and $Im[n] < 0$ for a Fano anti-laser. As described in [34], this structure can be analyzed through temporal CMT [36], following the set of equations (Fig. 1b)

$$\frac{da}{dt} = (i\omega_A - \frac{2}{\tau_w} - \frac{1}{\tau_s})a + \sqrt{\frac{2}{\tau_w}}(S_{+1} + S_{+2}) + \sqrt{\frac{2}{\tau_s}}S_{+s}, \quad (2)$$

$$S_{-1,2} = -S_{+1,2} + \sqrt{\frac{2}{\tau_w}}a, \quad S_{-s} = -S_{+s} + \sqrt{\frac{2}{\tau_s}}a, \quad S_{+s} = S_{-s}e^{-2ik_0nd},$$

where $\tau_w$ (or $\tau_s$) is the lifetime of the resonator coupled to an input/output waveguide (or stub), $\omega_A$ and $a$ are the resonant frequency and field of the junction resonator, respectively, $d$ is the stub length, and $k_0 = 2\pi/\lambda_0$ is the free-space wavenumber. Equation (2) derives $t(\omega)$ in Eq. (1) as

$$t(\omega) = \frac{\sqrt{\frac{2}{\tau_w}}}{i(\omega - \omega_a + \frac{2}{\tau_w} + \frac{1}{\tau_s}) - \frac{1}{\tau_s}[1 - i\tan(nk_0d)]}, \quad (3)$$

and $r_{11}(\omega) = r_{22}(\omega) = -1 + t(\omega)$. Note that because $D(\omega) = r_{11}(\omega)\cdot r_{22}(\omega) - t^2(\omega) = 1 - 2t(\omega)$, the singularity parameter $D(\omega)$ has a spectral response identical to that of $t(\omega)$.

Achieving high-Q perfect absorption with $D(\omega) \sim 0$ of a singularity can be realized with the control of the Fano parameter. We first investigate a lossless design and single port excitation for the CMT model ($Im[n] = 0$ and $S_{+2} = 0$, Fig. 1c,1d), for transmission ($T(\omega) = |t(\omega)|^2$) and reflection ($R(\omega) = |r(\omega)|^2$). Whereas the junction resonator connected to the waveguide is operated as a broadband continuum (dotted lines in Fig. 1c,1d), the stub plays a role of a narrowband discrete state due to the lack of a scattering path. According to the spectral position of the junction resonance $\omega_A$ relative to that of the stub resonance $\omega_0 = (2m+1)\pi c/(2\cdot Re[n]\cdot d)$, the degree of Fano asymmetry can be controlled (Fig. 1c vs 1d). Note that the asymmetry originates from the interference between broad and narrow phase information because the shifted resonance results in constructive and destructive interference for each side around the resonant frequency $\omega_0$ of the discrete state (Fig. 1e: destructive for both sides, Fig. 1f: constructive for $\omega > \omega_0$ and drastic destructive for $\omega < \omega_0$).

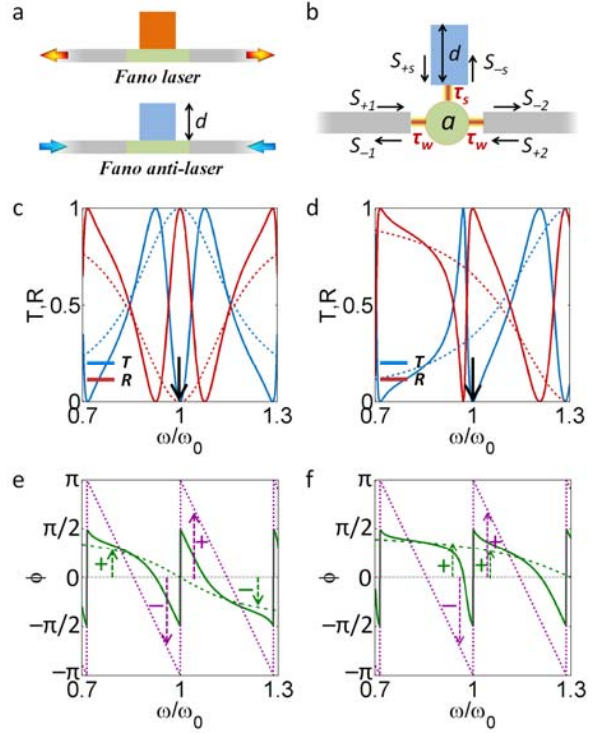

FIG. 1 (color online). Fano anti-laser system based on a stub-waveguide structure. (a) Schematic of Fano anti-laser compared to Fano laser dynamics (Orange for gain and blue for lossy stub, gray for feeding waveguides, and green for the junction). (b) CMT model of the structure in (a). Transmission ($T=|S_{-2}/S_{+1}|^2$, blue solid) and reflection ($R=|S_{-1}/S_{+1}|^2$, red solid) spectra for (c) non-Fano ($\omega_A = \omega_0$) and (d) Fano ($\omega_A \neq \omega_0$) cases, both with a single port excitation ($S_{+1}=1$, $S_{+2}=0$). Arrows in (c,d) denote the stub resonance ($\omega_0$), and dashed lines denote junction resonator responses without a stub ($\tau_s \rightarrow \infty$). The phase information for the non-Fano (c) and Fano (d) cases are shown in (e) and (f), respectively. Dotted lines denote the phases (purple for the stub phase -$2nk_0d$ and green for the junction phase $arg(a)$) before the connection between the stub and the junction, and solid green line shows the phase $arg(a)$ after the coupling. The stub length $d = (7/4)\cdot\lambda_0/Re[n]$ where $n = 1.5$. $\omega_A = 1.3\cdot\omega_0$ for the Fano case. $Q_w = \omega_A\tau_w/2 = 6$ and $Q_s = \omega_A\tau_s/2 = 6$.

Understanding the single port transmission $t(\omega)$ of the system, we now consider the absorption in the anti-lasing system (Fig. 2), first without Fano spectral asymmetry (Fig. 2a,2b, $\omega_A = \omega_0$). Figure 2c-2e shows the absorption with a single port incidence (Fig. 2a) for different values of coupling to the stub ($Q_s = \omega_A\tau_s/2$). The increase of $Q_s$ derives the enhancement of the absorption Q-factor $Q_{abs}$, whereas the peak value of the absorption is limited to 50% due to the existence of scattering (balanced absorption and scattering). However, following the physics of CPA, another 'coherent' incidence to the resonator (Fig. 2f,2g) can be introduced to completely suppress the scattering through destructive interference (Fig. 2h-2j), to stress, at the condition of $D(\omega) = 0$. Although the magnitude of absorption increases from 50% toward the limit of perfect absorption (100%), the bandwidth solely determined by $D(\omega)$ still remains identical for both cases. To realize high-$Q$ perfect absorption, a drastic spectral variation of $D(\omega)$ should be achieved, as inspired by the Fano system in Fig. 1, of the Fano anti-laser.

respectively, as functions of $Q_s$. An order of magnitude increase in $Q_{abs}$ is obtained in Fano anti-lasers (Fig. 3j, e.g., $Q_{abs}$ = 114 with $\omega_A = 1.5·\omega_0$ vs $Q_{abs}$ = 9 with $\omega_A = 1.0·\omega_0$ at $Q_s$ = 12) following the reduction of material loss (Fig. 3k). This result demonstrates that Fano resonance satisfies the singularity $D(\omega) = 0$ with considerably smaller loss from the phase-induced mirror, as studied intensively in the fields of dynamic $Q$ control [37] or bound states in the continuum (BIC) [38]. Furthermore, although only the bandwidth of the scattering has been treated in those works [37,38], we provided the method of controlling the absorption bandwidth for the first time from the concept of the Fano anti-laser.

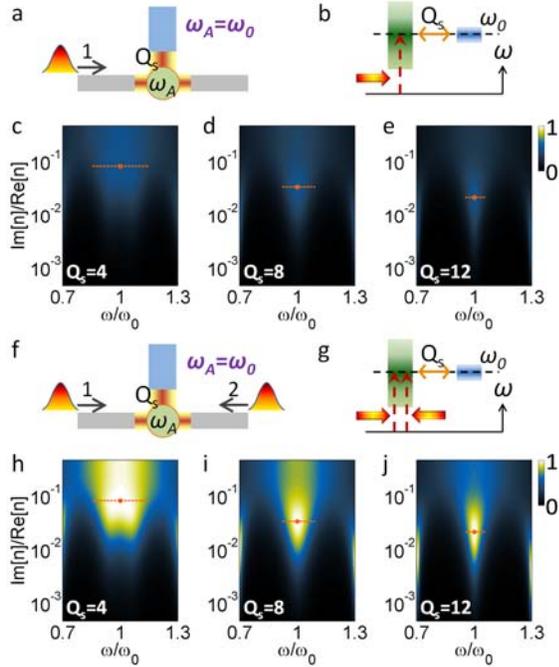

FIG. 2 (color online) Absorption spectra of a non-Fano anti-laser ($\omega_A = \omega_0$) with (a-e) a single port incidence and (f-j) coherent incidences, for (c,h) $Q_s$ = 4, (d,i) $Q_s$ = 8, and (e,j) $Q_s$ = 12. (a,f) Schematics and (b,g) energy level diagrams are shown for each case. The absorption spectra are obtained as a function of $Im[n] / Re[n]$. $d = (7/4)·(\lambda_0)/Re[n]$, and $Q_w = 6$.

As shown in Fig. 1d and [34], Fano asymmetry is achieved with the condition of $\omega_A \neq \omega_0$ (Fig. 3a,3b). Figures 3c-3e denote the absorption spectra of Fano-resonant systems, with $\omega_A = 1.3·\omega_0$ and coherent incidences. Compared to non-Fano systems (Fig. 2h-2j, e.g., $Q_{abs}$ = 9 with $Q_s$ = 12), the increase in $Q_{abs}$ is evident for each case of equal $Q_s$ (e.g., $Q_{abs}$ = 34 with $Q_s$ = 12) while preserving the condition of perfect absorption (100%). This high-$Q$ perfect absorption is also elucidated with the Fano-induced drastic spectral responses of $D(\omega)$ (Fig. 3f vs Fig. 3g,3h). With stronger asymmetry ($|\omega_A—\omega_0|\uparrow$), the bandwidth of the Fano anti-laser around the singularity ($D(\omega) = 0$) decreases significantly (Fig. 3i), in line with the transmission spectrum (Fig. 1d). Figures 3j and 3k show $Q_{abs}$ and required material parameters of $Im[n] / Re[n]$,

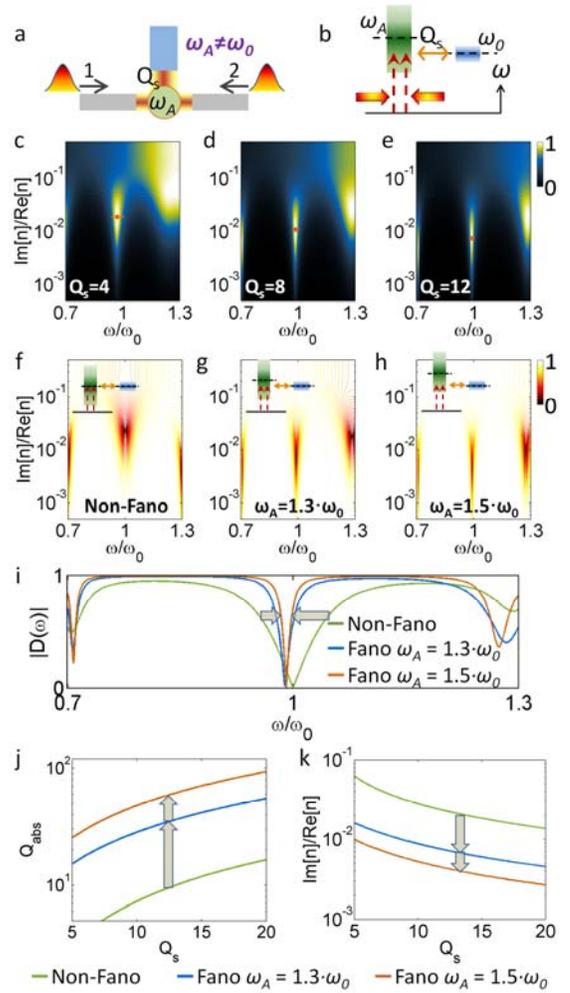

FIG. 3 (color online) (a-e) Absorption spectra of Fano anti-lasers ($\omega_A = 1.3·\omega_0$) with coherent incidences, for (c) $Q_s$ = 4, (d) $Q_s$ = 8, and (e) $Q_s$ = 12. (a) Schematics and (b) energy level diagram are also shown. (f-h) show $D(\omega)$ for (f) the non-Fano ($\omega_A = \omega_0$) and Fano cases of (g) $\omega_A = 1.3·\omega_0$ and (h) $\omega_A = 1.5·\omega_0$. $Q_s$ = 12 for (f-h). (i) Spectral variations of $D(\omega)$ for (f-h) at the perfect absorbing point of $Im[n]/Re[n]$. Variations of (j) $Q_{abs}$ and (k) required loss ($Im[n]/Re[n]$) for different degrees of Fano asymmetry are shown as functions of $Q_s$. Arrows in (i-k) present changes per the increase of Fano asymmetry. All other parameters are equal to those in Fig. 2.

To show the feasibility of high-$Q$ perfect absorption in a realistic platform, we employ a dielectric photonic crystal (PC, [39]): 2D square lattice of square rods (Fig. 4a, side length of rods $l = 0.4a$ and period $a = 500$nm). Waveguides, junction, and stub consist of defects ($l_d = 0.15a$), and the coupling between elements is controlled by the number of barrier rods. Note that the configuration of Fano resonances is realized by detuning each length of the stub $d_s$ and junction $d_A$. Figure 4b shows single port responses for different degrees of Fano asymmetry. The corresponding CPA spectra (Fig. 4c-4e) are in line with the results of CMT analysis; reduced material loss at the perfect absorption peaks (red dots) and high-$Q$ CPA with strong asymmetry are evident.

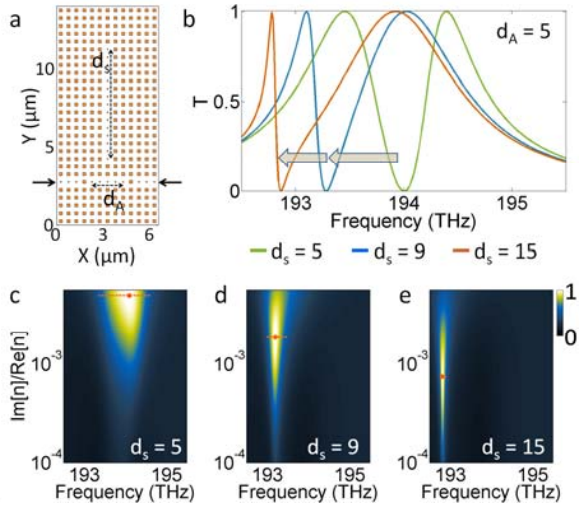

FIG. 4 (color online) (a) Fano anti-laser in PC realization ($d_A = 5$, $d_s = 15$). (b) Transmission spectra for a single port incidence. (c-e) Absorption spectra of PC Fano anti-lasers ($d_A = 5$) with (c) $d_s = 5$, (d) $d_s = 9$, and (e) $d_s = 15$. $Re[n] = 3.5$ for dielectric rods embedded in air. All the results are obtained using COMSOL.

It is noted that the high-$Q$ absorption of PC Fano anti-laser well satisfies the requirement of selective thermal emitters [4,15,16]. Figures 5a-5f show sharp change of field distribution (Fig. 5a-5c) at the Fano-resonance frequency, deriving selective resistive heating of the stub (Fig. 5d-5f). As seen, strong heating occurs only inside the stub of a discrete state, selectively at the perfect absorbing frequency (Fig. 5b,5e). These results demonstrate that the spatial distribution of energy concentration (intensity enhancement: 110 in the stub and 9 in the junction at 192.8THz) determines the heat generation map of the Fano anti-lasing system: allowing not only the spectral separation ($\omega_0$ vs $\omega_A$) but also the spatial separation (stub vs junction) in opto-thermal conversion, by the spatial distribution of high-$Q$ (stub) and low-$Q$ (junction) elements with different resonant frequencies. Not restricted to thermal emitters [4,15,16], we note that such a spatio-spectral selectivity is also desired in hyperspectral thermal imaging [14]: in our design, spectrally-selective imaging of a spatially discrete state of a stub only.

The reduction of the required material loss can also be utilized for absorptive switching, where the switching power is proportional to $Im[n]$ [6]. Figure 5g shows the drastic decrease of the required $Im[n]$, for the on-off switching at the condition of perfect absorption. The maximum modulation depth (0-100%) is enabled by the CPA, and the reduction in the required power, by an order of magnitude, is achieved by aligning the resonance of the stub (with $d_s$) to the point of extreme Fano asymmetry. This property allows the Fano anti-laser an attractive platform, especially for absorptive switching. We note that the proposed Fano switching in CPA can also be achieved by controlling the coherence condition [40] or real optical potential $Re[n]$ [41]. In terms of the real implementation, the difficulty in the realization of rod-type photonic crystals, which mainly originates from the high aspect ratio of rods ($l / h \sim 5$ where $h$ is the height of rods [39]), can be overcome by applying triangular-lattice hole-type photonic crystals ($l_{hole} / h \sim 0.8$ where $l_{hole}$ is the diameter of holes [39]).

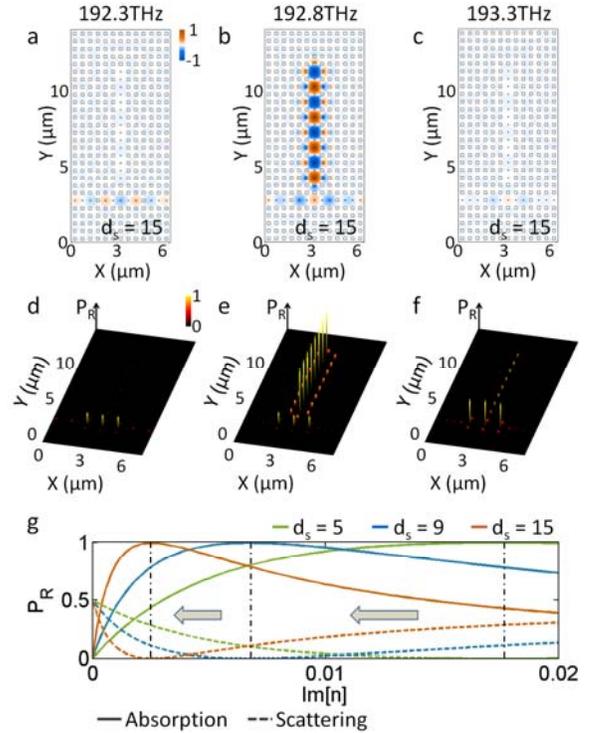

FIG. 5 (color online) (a-c) Field distributions and (d-f) heat generation maps around the perfect absorption (b,e), for selective thermal emitters ($d_A = 5$, $d_s = 15$). (g) Switching as a function of $Im[n]$ for different degrees of Fano spectral asymmetry. Solid (or dotted) lines denote the absorbed power (or the scattering for a single port). All other parameters are equal to those in Fig. 4.

In conclusion, by using the time-reversed form of a Fano laser, we demonstrated the high-$Q$ perfect absorption of the Fano anti-laser; to overcome the absence of the anti-lasing state above the threshold. The inherent tradeoff between large absorption and narrow bandwidth was successfully overcome. The spectral singularity parameter $D(\omega)$ from CMT offers the condition of perfect absorption, and also reveals the origin of the drastic spectral response, i.e. Fano asymmetry. A highly sensitive singularity in scattering, which derives an order of increasing $Q$ for CPA spectra, is achieved. In terms of application, the Fano anti-laser using phase-dependent reflection is in a similar vein as dynamic $Q$ devices [37] or BIC [38], but in the context of manipulating 'absorption' spectra, extends the boundary of high-$Q$ optical element which was limited to scattering spectrum manipulation.

Although we focused on the realization in a guided-wave platform, both the S-matrix singularity condition for perfect absorption and the coupled mode analysis for Fano resonances [42] can also be satisfied in other general platforms, including the structure for free-space waves. Our results can thus be extended to other generalized Fano-resonant platforms [42], including ring resonators and grating structures. The Fano anti-laser paves the way toward 'complete' energy conversion from optical to internal energy yet retaining high-$Q$ spectral operation, and we envisage its applications based on the 'single-wavelength absorber' in switches [6,40,41], sensors [1-3,12,13], and opto-thermal emitters [4,15,16].

This work was supported by the National Research Foundation through the Global Frontier Program (GFP) NRF-2014M3A6B3063708, the Global Research Laboratory (GRL) Program K20815000003, and the Brain Korea 21 Plus Project in 2015, all funded by the Ministry of Science, ICT & Future Planning of the Korean government.

*nkpark@snu.ac.kr
†These authors contributed equally to this work.